# Phase coherent excitation of SABRE permits simultaneous hyperpolarization of multiple targets at high magnetic field

Jacob R. Lindale[a], Shannon L. Eriksson[a,b], Warren S. Warren[c]

[a]Department of Chemistry, Duke University, Durham, NC, 27708. [b] School of Medicine, Duke University, Durham, NC, 27708. [c]Departments of Physics, Biomedical Engineering, and Radiology, Duke University, Durham, NC, 27708.

Hyperpolarization methods in magnetic resonance overcome sensitivity limitations, especially for low-$\gamma$ nuclei such as $^{13}$C and $^{15}$N. Signal Amplification By Reversible Exchange (SABRE) and extended SABRE (X-SABRE) are efficient and low-cost methods for generating large polarizations on a variety of nuclei, but they most commonly use low magnetic fields ($\mu$T-mT). High field approaches, where hyperpolarization is generated directly in the spectrometer, are potentially much more convenient but have been limited to selectively hyperpolarize single targets. Here we introduce a new pulse sequence-based approach that affords broadband excitation of SABRE hyperpolarization at high magnetic fields without having to tailor pulse sequence parameters to specific targets. This permits simultaneous hyperpolarization of multiple targets for the first time at high field and offers a direct approach to integration of high-field SABRE hyperpolarization into routine NMR applications, such as NMR-based metabonomics and biomolecular NMR.

## Introduction

Nuclear Magnetic Resonance (NMR) spectroscopy is a powerful and versatile characterization technique. However, it is limited by low signal intensities, which worsen for low-$\gamma$ nuclei like $^{13}$C and $^{15}$N due to low spin polarization along the leading magnetic field. For applications such as biomolecular NMR, this largely limits heteronuclear spectroscopy to time-intensive multidimensional correlation experiments using proton ($^{1}$H) detection. In metabonomics, the inherently low NMR signal intensity also restricts conventional use to $^{1}$H spectra, which for complex samples like biofluids are highly congested due to the low chemical shift dispersion of $^{1}$H relative to other nuclei.

Hyperpolarization methods(1-4) circumvent these issues by distilling large, non-equilibrium polarization from an external source of spin order. This results in enhanced signals that are observable in single-scan NMR spectra, reducing acquisition times and lowering detection thresholds. Over the last decade, the hyperpolarization method Signal Amplification By Reversible Exchange (SABRE)(4) has attracted much attention, as it is inexpensive and can generate high polarizations in under a minute(5) (Figure 1A). While the original demonstrations of SABRE targeted $^{1}$H(4, 6), later variants extended SABRE to heteronuclei(7-13), which exhibit superior relaxation lifetimes and spectral dispersion relative to $^{1}$H (these variants are generically called extended

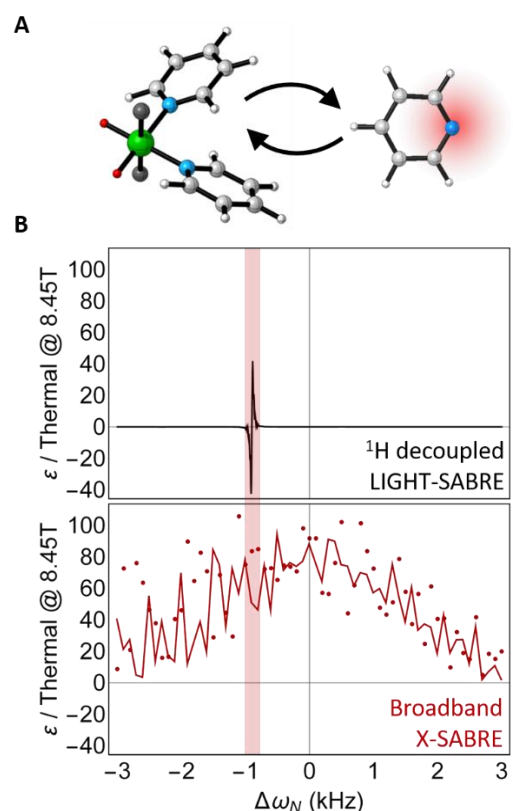

**Fig. 1 Top:** In SABRE and X-SABRE hyperpolarization methods, a polarization transfer catalyst reversibly binds both para-hydrogen and a molecular target (here shown as pyridine) and singlet order is converted into hyperpolarized target magnetization on the target. Exchanging back into solution forms hyperpolarized free species and polarizes more molecules. **Bottom:** In previous methods such as $^{1}$H decoupled LIGHT-SABRE, the hyperpolarization is created by weak irradiation of the iridium-bound heteronucleus, leading to narrow excitation bandwidths. Here, we demonstrate a broadband and phase-coherent excitation strategy that vastly increases hyperpolarization bandwidth. In the broadband case, experimental data are shown with points and simulation is shown by the line.

SABRE or X-SABRE). Hyperpolarization is sourced from parahydrogen (*p*-H$_2$), the singlet spin isomer of H$_2$, during transient interactions mediated by an iridium catalyst. Binding to the catalyst establishes *J*-couplings between the parahydrogen derived hydrides and target nuclei, through which polarization can flow "spontaneously" at ultralow magnetic fields(8) or using radiofrequency pulses(7, 10) in a conventional NMR spectrometer. Parahydrogen can be generated in large quantities and is relatively inexpensive to add to any NMR system.

High field X-SABRE methods are attractive, as no specialized equipment is required in addition to the parahydrogen apparatus. The original high field X-SABRE experiments, LIGHT-SABRE(7) and RF-SABRE(12), use weak radiofrequency irradiation to transfer spin order from the parahydrogen singlet state into magnetization on the heteronucleus. More recent work has led to a diverse set of high field X-SABRE variants that augment and expand the capabilities of the original experiments(14-20) (Table 1). All of these experiments were designed to generate hyperpolarization in systems with only a single ligand present in solution. The parahydrogen derived hydrides have the same chemical shift when the same ligand occupies both trans-hydride positions. However, most SABRE targets require a coligand to enhance exchange with the polarization transfer catalyst.

Multiple ligands lead to environments where the *p*-H$_2$ derived hydrides have different chemical shifts. Adding $^1$H-decoupling(21) to the LIGHT-SABRE pulse sequence permitted hyperpolarization in these asymmetric ligand environments and broadened the scope of high field X-SABRE. With the exception of the original INEPT-SABRE(15) experiment, all high field methods use selective excitation and require prior knowledge of the chemical shift of the iridium-bound heteronucleus to work. The narrow excitation bandwidth of these sequences is not conducive to searching for the X-SABRE resonance condition (Figure 1B). While the INEPT-SABRE(15) experiment has a larger bandwidth, this experiment can only transfer hyperpolarization to one ligand per catalyst.

In this paper, we overcome these limitations by introducing a pulse sequence that generates phase-coherent excitation of hyperpolarized SABRE magnetization. Phase-coherent excitation provides more control over the system evolution and is used here to compensate for unwanted modulation in the hyperpolarization when irradiating off-resonance. This increases the excitation bandwidth over previous experiments (**Fig. 1**) and facilitates simultaneous hyperpolarization of multiple targets at high magnetic field for the first time. While X-SABRE hyperpolarization of multiple targets has been demonstrated at ultralow magnetic field(22), the absence of the chemical shift interaction at these fields eliminates chemical specificity. Hyperpolarization of multicomponent systems at high field can be done without any significant alteration to the pulse sequence, making it highly robust to changes in hyperpolarization target.

| | Experiment | Symmetric ligand environments | Asymmetric ligand environments | Coherently pump polarization | Catalytic repolarization | Insensitive to changing target | Multiple targets simultaneously |
|---|---|---|---|---|---|---|---|
| **Ultralow Field** | SABRE-SHEATH (8) | ✓ | ✓ | ✗ | ✓ | ✓ | ✓ |
| | Coherent SHEATH (20) | ✓ | ✓ | ✓ | ✓ | ✓ | ✓ |
| **High Field** | LIGHT-SABRE (7) | ✓ | ✗ | ✗ | ✓ | ✗ | ✗ |
| | RF-SABRE (12) | ✓ | ✗ | ✗ | ✓ | ✗ | ✗ |
| | INEPT-SABRE (15) | ✓ | ✗ | ✓ | ✗ | ✓ | ✗ |
| | Re-INEPT-SABRE (14) | ✓ | ✗ | ✓ | ✓ | ✗ | ✗ |
| | ADAPT-SABRE (18) | ✓ | ✗ | ✗ | ✓ | ✗ | ✗ |
| | SLIC-SABRE (19) | ✓ | ✗ | ✗ | ✓ | ✗ | ✗ |
| | DARTH-SABRE (20) | ✓ | ✗ | ✓ | ✓ | ✗ | ✗ |
| | $^1$H dLIGHT-SABRE (21) | ✓ | ✓ | ✓ | ✓ | ✗ | ✗ |
| | **Broadband X-SABRE** | ✓ | ✓ | ✓ | ✓ | ✓ | ✓ |

**Table 1** Comparison of X-SABRE experiments and scope of performance. The ultralow field SABRE-SHEATH experiment and its coherently pumped variant are versatile experiments that are robust to a variety of experimental conditions. The broadband X-SABRE experiment presented here (bold) augment previous advances of high field experiments to exhibit comparably robust performance to the ultralow field variants.

## Theory

SABRE generates hyperpolarization by converting the singlet order of *p*-H$_2$ into observable magnetization (or other spin orders) on the target nucleus. Only a small fraction of the hyperpolarization target is bound to the SABRE catalyst at any time, so the transfer of spin order must occur repeatedly to maximize the number of targets onto which spin order is transferred and thus maximize the resulting polarization.

Singlet order transfer in high field SABRE is often visualized as a level anti-crossing (LAC) in a rotating frame, where population is allowed to transfer between the initial singlet state and a magnetized state. This has been realized using variations of Spin Lock Induced Crossing (SLIC) pulses(7, 10-12, 19, 20, 23). The desired LAC condition is generated when the pulse power meets the resonance condition $\omega_{1,N} \approx J_{HH} \approx 10\ Hz$, where $J_{HH}$ is the *J*-coupling between the parahydrogen-derived hydrides. The weak irradiation required to satisfy the LAC condition makes SLIC-based experiments naturally selective. Coherence transfer methods (INEPT-SABRE(14, 15) or ADAPT-SABRE(18)) have also used selective $^{15}$N-irradiation so that multiple applications of these sequence may be applied. Selectivity has the benefit that the pulse does not affect the polarization accumulated on the hyperpolarized target after it exchanges off of the complex. However, it requires that the chemical shift of the iridium-bound heteronucleus is known prior to the experiment, and has limited the scope of high field X-SABRE applications.

Herein, our objective is to expand the scope of high field X-SABRE to match that of the ultralow field variants. will examine the ABX system for X = $^{15}$N as a model system for $^{15}$N-SABRE. As hyperpolarization in SABRE is generated on the iridium catalyst, we will focus on this species, referred to as the 'bound species'. The ABX Hamiltonian is ($\hbar = 1$):

$$\widehat{\mathcal{H}}_0 = \omega_{0,1}\hat{I}_{1z} + \omega_{0,2}\hat{I}_{2z} + \omega_{0,N}\hat{S}_z \\ + 2\pi\big(J_{HH}\hat{I}_1 \cdot \hat{I}_2 + J_{NH}\hat{I}_{1z}\hat{S}_z\big) \quad (1)$$

In this notation, $\hat{I}$ are the spin operators corresponding to the hydrides and $\hat{S}$ corresponds to the heteronucleus. The first three terms correspond to the nuclear Zeeman and chemical shift interactions. The *J*-coupling term $\hat{I}_1 \cdot \hat{I}_2$ is the hydride-hydride coupling and $\hat{I}_{1z}\hat{S}_z$ is the heteronuclear $^1$H-$^{15}$N *J*-coupling. Importantly, the transverse terms of the heteronuclear coupling are negligible due to the large resonance frequency difference between these nuclei. Transforming this Hamiltonian into a rotating frame simplifies the discussion of the dynamics. In this case, the $\hat{I}$ spins will be rotated about the center frequency of the hydrides and the $\hat{S}$ spin will be rotated about the $^{15}$N pulse carrier frequency ($\omega_{rf}$), giving:

$$\widehat{\mathcal{H}}_0 = \Delta\omega_{HH}\big(\hat{I}_{1z} - \hat{I}_{2z}\big) + \Delta\omega_N\hat{S}_z \\ + \omega_{1,H}\big(\hat{I}_{1x} + \hat{I}_{2x}\big) + \omega_{1,N}\hat{S}_x \quad (2) \\ + 2\pi\big(J_{HH}\hat{I}_1 \cdot \hat{I}_2 + J_{NH}\hat{I}_{1z}\hat{S}_z\big)$$

We have used $\Delta\omega_{HH} = \omega_{0,1} - \omega_{0,2}$ and $\Delta\omega_N = \omega_{0,N} - \omega_{rf,N}$, and it is assumed that the $^1$H pulse carrier frequency is tuned to $\omega_{rf,H} = (\omega_{0,1} + \omega_{0,2})/2$ for brevity.

The initial state of a SABRE experiment can be written as the tensor product between the parahydrogen singlet state and thermal polarization on $^{15}$N, which is assumed to be small enough to be simply proportional to unity ($\hat{E}$), giving:

$$\hat{\rho}_0 = \left(\frac{1}{4}\hat{E} - \hat{I}_1 \cdot \hat{I}_2\right) \otimes \left(\frac{\hat{E}}{2}\right) \quad (3)$$

The only important term here is $\hat{I}_1 \cdot \hat{I}_2$, the non-unity component of the singlet state, which is the form that was assumed in the initial treatments of SABRE. However, $\Delta\omega_{HH}$ is much larger than all other terms in

$\hat{\mathcal{H}}_0$ at 8.45T and the transverse terms of $\hat{I}_1 \cdot \hat{I}_2$ accumulate a phase as:

$$e^{-i\Delta\omega_{HH}(\hat{I}_{1z}-\hat{I}_{2z})t}\hat{I}_1 \cdot \hat{I}_2 e^{i\Delta\omega_{HH}(\hat{I}_{1z}-\hat{I}_{2z})t}$$
$$= \cos(\Delta\omega_{HH}t)\left(\hat{I}_{1x}\hat{I}_{2x} + \hat{I}_{1y}\hat{I}_{2y}\right) \quad (4)$$
$$+ \sin(\Delta\omega_{HH}t)\left(\hat{I}_{1x}\hat{I}_{2y} - \hat{I}_{1y}\hat{I}_{2x}\right) + \hat{I}_{1z}\hat{I}_{2z}$$

As *p*-H$_2$ exchanges with the catalyst, the evolution described in **eq. 4** is shifted in time and results in the loss of phase coherence for the transverse terms. A similar dephasing is generated by evolution under the $J_{NH}$ coupling in symmetric systems. Therefore, the effective initial state of the system is:

$$\hat{\rho}_0 = \left(\frac{1}{4}\hat{E} - \hat{I}_{1z}\hat{I}_{2z}\right) \otimes \left(\frac{\hat{E}}{2}\right) \quad (5)$$

However, SLIC pulses are only capable of converting the transverse terms of the singlet order ($\hat{I}_{1x}\hat{I}_{2x} + \hat{I}_{1y}\hat{I}_{2y}$) into heteronuclear magnetization. Thus, hyperpolarization is only generated using a SLIC pulse for symmetric environments, where these terms are partially preserved. The SLIC-SABRE experiment(19) showed that converting the initial $\hat{I}_{1z}\hat{I}_{2z}$ spin order into $\hat{I}_{1x}\hat{I}_{2x}$ using a $90_y$ pulse was possible in symmetric systems(19) and results in higher polarization levels compared to the LIGHT-SABRE experiment. In asymmetric systems(21), it was shown that strong $^1$H decoupling could effectively remove the $\Delta\omega_{HH}$ term and spin-lock the singlet state in $\hat{I}_1 \cdot \hat{I}_2$. However, this method requires strong and continuous $^1$H irradiation even at modest (8.45T) magnetic fields, preventing translation to higher fields required for applications such as biomolecular NMR.

When considering the design of a new X-SABRE experiment, it would be ideal to utilize the $\hat{I}_{1z}\hat{I}_{2z}$ term as it commutes with the other terms in **eq. 2** and is protected from evolution in the absence of pulses. This allows for the accumulation of the $\hat{I}_{1z}\hat{I}_{2z}$ spin order when hyperpolarization is not being transferred to the heteronucleus. Additionally, the experiment should be designed to generate broadband excitation of SABRE hyperpolarization. Finally, magnetization on the free species should be unaffected by the pulse sequence within the excitation bandwidth.

The analysis of coherence pathways can provide insight on the design of new X-SABRE experiments. The density matrix evolves according to the Liouville-von Neumann equation:

$$\frac{\partial}{\partial t}\hat{\rho} = i[\hat{\rho}, \hat{\mathcal{H}}] \quad (6.1)$$

$$\hat{\rho}(t) = \vec{\mathcal{T}}e^{-i\int_0^t dt' \hat{\mathcal{H}}(t')}\hat{\rho}_0 e^{i\int_0^t dt' \hat{\mathcal{H}}(t')} \quad (6.2)$$

$\vec{\mathcal{T}}$ is the Dyson time-ordering operator and the exponential is often called the propagator, $\hat{U}$. Coherence pathways can be found by using the Taylor expansion of $\hat{\rho}(t)$:

$$\hat{\rho}(t) = \sum_{n=0}^{\infty}\frac{t^n}{n!}\frac{\partial^n}{\partial t^n}\hat{\rho}(t) = \sum_{n=0}^{\infty}\hat{\rho}_n(t)$$
$$= \hat{\rho}_0 + i[\hat{\rho}_0, \hat{\mathcal{H}}]t - [[\hat{\rho}_0, \hat{\mathcal{H}}], \hat{\mathcal{H}}]\frac{t^2}{2} + \cdots$$
$$= \hat{\rho}_0 + i[\hat{\rho}_0, \hat{\mathcal{H}}]t + i\left[\frac{\partial}{\partial t}\hat{\rho}, \hat{\mathcal{H}}\right]\frac{t^2}{2} + \cdots \quad (7)$$

Here, we use $\hat{\rho}^n(t)$ as the $n^{th}$ term in the expansion. For non-exchanging systems, the maximum length of the coherence pathway (the number of transformations that can be applied to a system before it loses coherence) is often dictated by the relaxation rates of the system. Coherence pathway lengths in SABRE are limited by exchange, as targets are transiently bound to the iridium complex for tens of milliseconds. Hence, it is important to search for the leading term in **eq. 7** that generates magnetization on the target ($\hat{S}_z$). The initial spin order, $\hat{I}_{1z}\hat{I}_{2z}$, commutes with the Hamiltonian (**eq. 2**) and thus all higher order terms beyond $\hat{\rho}_0$ vanish.

To generate hyperpolarization, the form of the Hamiltonian must change so that it no longer commutes with the initial spin state. Waugh and co-workers realized that the *effective* form of the Hamiltonian may be altered by using radiofrequency pulses(24, 25). As a simple example, consider a sequence of two pulses ($\hat{U}_1, \hat{U}_2$), each followed by periods where the static Hamiltonian is active ($e^{i\hat{\mathcal{H}}_0 t_1}, e^{i\hat{\mathcal{H}}_0 t_2}$). The final density matrix is

$$\hat{\rho}(t) = \{e^{-i\hat{\mathcal{H}}_0 t_2}\hat{U}_2 e^{-i\hat{\mathcal{H}}_0 t_1}\hat{U}_1\}\hat{\rho}_0\{\cdots\}^\dagger \quad (8)$$

where the terms to the right-hand side of $\hat{\rho}_0$ have been truncated for brevity, but are the conjugate transpose of all terms to the left-hand side of $\hat{\rho}_0$. We can add identity $(\widehat{U}_1 \widehat{U}_1^\dagger)$ as

$$\begin{aligned}
\hat{\rho}(t) &= \{e^{-i\widehat{\mathcal{H}}_0 t_2} \widehat{U}_2 (\widehat{U}_1 \widehat{U}_1^\dagger) e^{-i\widehat{\mathcal{H}}_0 t_1} \widehat{U}_1\} \hat{\rho}_0 \{\cdots\}^\dagger \\
&= \{e^{-i\widehat{\mathcal{H}}_0 t_2} \widehat{U}_2 \widehat{U}_1 (\widehat{U}_1^\dagger e^{-i\widehat{\mathcal{H}}_0 t_1} \widehat{U}_1)\} \hat{\rho}_0 \{\cdots\}^\dagger
\end{aligned} \quad (9)$$

If we additionally assume that in this case $\widehat{U}_2 \widehat{U}_1 = \widehat{E}$, such as if the pulse flip angles sum to a multiple of $2\pi$, we may write:

$$\hat{\rho}(t) = \{e^{-i\widehat{\mathcal{H}}_0 t_2} (\widehat{U}_1^\dagger e^{-i\widehat{\mathcal{H}}_0 t_1} \widehat{U}_1)\} \hat{\rho}_0 \{\cdots\}^\dagger \quad (10)$$

We can rewrite the term in brackets as:

$$\widehat{U}_1^\dagger e^{-i\widehat{\mathcal{H}}_0 t_1} \widehat{U}_1 = e^{-i \widehat{U}_1^\dagger \widehat{\mathcal{H}}_0 \widehat{U}_1 t_1} \quad (11)$$

This appears as if the system is evolving under the Hamiltonian $\widehat{U}_1^\dagger \widehat{\mathcal{H}}_0 \widehat{U}_1$ during $t_1$ and indicates that the form of $\widehat{\mathcal{H}}_0$ during this period can effectively altered using the pulses proceeding $t_1$. This is generally referred to as the toggling frame transformation.

While the toggling frame indicates how pulses can act on $\widehat{\mathcal{H}}_0$ to alter its form, the Hamiltonian for the pulse sequence remains time-dependent, preventing straight-forward interpretation of the pulse sequence. Instead, the effective action of the time-dependent Hamiltonian may be calculated using average Hamiltonian theory, where the terms are given by the Magnus expansion. The first term is simply the time-average of $\widehat{\mathcal{H}}(t)$:

$$\overline{\mathcal{H}}^{(0)} = \frac{1}{T} \int_0^T dt\, \widehat{\mathcal{H}}(t) \quad (12)$$

Higher order corrections to the average Hamiltonian are possible, but will only be important when $T$ is long relative to the evolution of the Hamiltonian.

We can now explore coherence pathways that are accessible not only using the interactions of $\widehat{\mathcal{H}}_0$ but also interactions that would be generated by rotations of $\widehat{\mathcal{H}}_0$. The interactions susceptible to rotation by pulses are the Zeeman terms ($\hat{I}_{1z} - \hat{I}_{2z}$ and $\hat{S}_z$) and the heteronuclear J-coupling ($\hat{I}_z \hat{S}_z$). Thus, terms like $\hat{I}_{1x} - \hat{I}_{2x}$ or $\hat{S}_x$ can be generated from the Zeeman interactions using a $90_y$ pulse and the J-coupling can produce terms like $\hat{I}_y \hat{S}_y$, if $90_x$ pulses are applied to both ¹H and ¹⁵N. The $J_{HH}$ coupling ($\hat{I}_1 \cdot \hat{I}_2$) is a zero-rank tensor and is thus rotationally invariant, leaving it unaffected by pulses. The shortest coherence pathway from the initial $\hat{I}_{1z} \hat{I}_{2z}$ spin order to magnetization on the target, $\hat{S}_z$, can then be found using terms from $\widehat{\mathcal{H}}_0$ and terms that may be accessed by rotations induced by pulses. The most efficient of these pathways generates magnetization in the third term of the Taylor expansion and has the form (all units in rad s⁻¹):

$$\begin{aligned}
\hat{\rho}^I(t) &= i[\hat{I}_{1z} \hat{I}_{2z}, \hat{I}_{1y} \hat{S}_y](J_{NH} t) \\
&= \hat{I}_{1x} \hat{I}_{2z} \hat{S}_y (J_{NH} t)
\end{aligned} \quad (13.1)$$

$$\begin{aligned}
\hat{\rho}^{II}(t) &= i[\hat{I}_{1x} \hat{I}_{2z} \hat{S}_y, \hat{I}_1 \cdot \hat{I}_2]\left(J_{NH} J_{HH} \frac{t^2}{2}\right) \\
&= \hat{I}_{1y} \hat{S}_y \left(\frac{J_{NH} J_{HH}}{4} \frac{t^2}{2}\right)
\end{aligned} \quad (13.2)$$

$$\begin{aligned}
\hat{\rho}^{III}(t) &= i[\hat{I}_{1y} \hat{S}_y, \hat{I}_{1y} \hat{S}_x]\left(\frac{J_{NH}^2 J_{HH}}{4} \frac{t^3}{6}\right) \\
&= \hat{S}_z \left(\frac{J_{NH}^2 J_{HH}}{16} \frac{t^3}{6}\right)
\end{aligned} \quad (13.3)$$

This coherence pathway involves the action of heteronuclear J-coupling terms that are not present in $\widehat{\mathcal{H}}_0$ but are available by rotations of $\widehat{\mathcal{H}}_0$ by pulses. Importantly, this coherence pathway is independent of the resonance offset of all of the spins as well as the power of the pulses used in excitation, unlike the case of the SLIC-based SABRE experiments (see SI). It should be noted that there are multiple pathways that generate polarization using different sets of phases of the heteronuclear J-coupling. Each of these pathways accumulates phase differently and present the possibility of using phase cycling methods.

The heteronuclear J-couplings required to excite the target coherence pathway can be generated by the pulse sequence in **Fig. 2**. Two 'excitation blocks' (delineated in the figure by color) are applied sequentially to generate the desired form of the heteronuclear J-coupling during each period. The delay $\tau$ allows for evolution under the rotated heteronuclear J-couplings and the period $\tau_{HH}$ permits evolution under $J_{HH}$. The phase of the first pulse in each block can be chosen to set the phase of that spin of the $J$-coupling. For example, the first $\bar{x}$ pulse on the $^{15}$N channel in **Fig. 2** generates a $\bar{x} - \pi/2 = y$ phase component of the coupling. The phase of each spin operator in the $J$-coupling may be independently changed by simply changing the pulse phase. Importantly, the term $\hat{I}_z \hat{S}_z$ is only present during $\tau_{HH}$ when it commutes with all of the spin operator terms present. The average Hamiltonians during each excitation block and $\tau_{HH}$ are:

$$\bar{\mathcal{H}}^{(0)}(t) = \begin{cases} 2\pi \begin{pmatrix} J_{HH}\hat{I}_1 \cdot \hat{I}_2 \\ +J_{NH}\hat{I}_{1y}\hat{S}_y \end{pmatrix} & 0 < t \leq t_1 \\ 2\pi \begin{pmatrix} J_{HH}\hat{I}_1 \cdot \hat{I}_2 \\ +J_{NH}\hat{I}_{1z}\hat{S}_z \\ +\hat{\mathcal{H}}_Z \end{pmatrix} & t_1 < t \leq t_2 \\ 2\pi \begin{pmatrix} J_{HH}\hat{I}_1 \cdot \hat{I}_2 \\ +J_{NH}\hat{I}_{1y}\hat{S}_x \end{pmatrix} & t_2 < t \leq t_3 \end{cases} \quad (14)$$

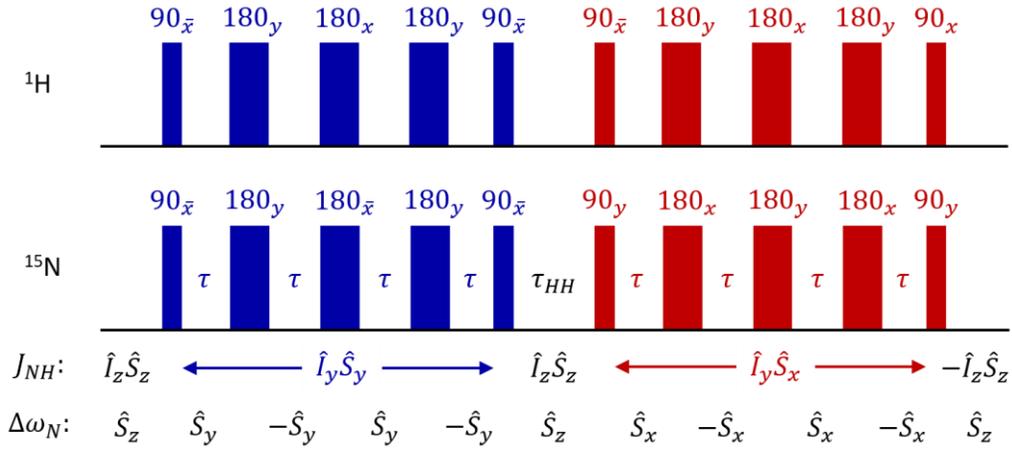

**Fig. 2** Pulse sequence required to transfer hyperpolarized singlet order into magnetization on the SABRE target. The $\tau$ delay permits evolution under $J_{NH}$ and the $\tau_{HH}$ delay permits evolution under $J_{HH}$. The blue and red excitation blocks delineate the formation of the two, phase shifted heteronuclear $J$-couplings in the average Hamiltonian. The first and third 180° pulses of each block refocus unwanted phase evolution only. The toggling frame is shown below the sequence. Importantly, the pulse flip angles sum to $2\pi n$ and thus generate no net excitation of the hyperpolarized free species. $\bar{x} = -x$.

In this notation, $\hat{\mathcal{H}}_Z$ is the Zeeman Hamiltonian and $t_1 = 4\tau$, $t_2 = 4\tau + \tau_{HH}$, and $t_3 = 8\tau + \tau_{HH}$. Keeping the Hamiltonian piecewise time-independent is convenient as it emphasizes the similarities between the constructed Hamiltonian and the chosen coherence pathway shown in **eq. 13**. Importantly, this Hamiltonian will be valid within the pulse excitation bandwidth. Note that each excitation block separately exerts a flip angle that is $2\pi n$ on both $^1$H and $^{15}$N. This means that it can be repeatedly applied directly to the free target resonance without destroying the accumulated hyperpolarization.

The three-spin coherence excited during the first excitation block evolves under $\hat{\mathcal{H}}_Z$. As such, the coherent dynamics during $\tau_{HH}$ oscillate depending on the resonance offset of the individual spins. The $\tau_{HH}$ delay would have to be carefully optimized for a given set of resonance offsets, significantly limiting applications. While it is simple enough to remove evolution under $\hat{\mathcal{H}}_Z$ with an extra 180° pulse, this would make the net flip angle of the pulse sequence $(2n + 1)\pi$, meaning that the polarization would invert with each sequential application and this leads to an attenuation of the hyperpolarization (see SI). Instead, the dependence of $\hat{\mathcal{H}}_Z$ can be removed by cycling the excitation phase of the two excitation blocks. For instance, when using the excitation phase $yy + yx$ shown in **Fig. 2**, the hyperpolarized signal will acquire a dependence proportional to $\cos(\Delta\omega_{HH}\tau_{HH}/2)\cos(\Delta\omega_N\tau_{HH})$. However, if the excitation phase was $xy + yx$, then the signal would be proportional to $\sin(\Delta\omega_{HH}\tau_{HH}/2)\cos(\Delta\omega_N\tau_{HH})$. Thus, the four-step phase cycle $\{yy, xy, yx, xx\} + yx$ can be used to eliminate dependence on $\hat{\mathcal{H}}_Z$, where separate experiments are collected for each of the phases in the first excitation block and linearly combined.

While the average Hamiltonian treatment would be sufficient for non-exchanging systems, chemical exchange must be considered to accurately predict and optimize the performance of the pulse sequence. The physically rigorous and experimentally validated SABRE-specific DMExFR2(26) model was used to simulate the performance of the pulse sequence under realistic conditions (**Fig. 3**). In this model, all of the relevant physical effects are included, such as coligand effects, exchange between stereoisomers of the catalyst, rebinding of previously polarized ligands, and relaxation. For these simulations, the parameters were modelled using the parameters of the $^{15}$N-acetonitrile SABRE system with a pyridine (natural abundance) coligand. At 8.45T, the relevant parameters are $\Delta\omega_{HH} = 460\ Hz$, $\Delta\omega_{N,b} = -900\ Hz$ (offset of bound species), $\Delta\omega_{N,f} = 900\ Hz$ (offset of free species), $J_{HH} = -8.68\ Hz$, and $J_{NH} = -25.71\ Hz$. For both $^{15}$N and $^{1}$H, rectangular pulses with powers of $\omega_1 = 2.5\ kHz$ (90°) or $5\ kHz$ (180°) were used. Additionally, the pulse sequence is applied 50 times using a delay of 400 ms between each sequence to allow for exchange of the polarized target off of the catalyst. Under these conditions, the $\tau$ delay is optimized to be approximately 3 ms, with finer structure appearing in the dynamics when the frequency difference is close to the excitation bandwidth (**Fig. 3A**). Simulating the four steps of the proposed phase cycle indicate that when linearly combined, hyperpolarization is obtained largely invariant to the precise timing of the $\tau_{HH}$ delay (**Fig. 3B**). Interestingly, residual oscillatory structure is also observed as a result of the resonances approaching the edge of the excitation bandwidth.

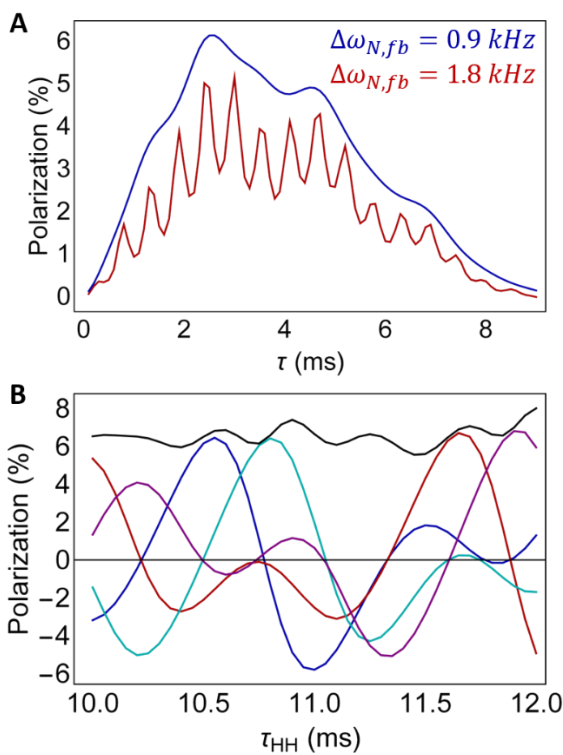

**Fig. 3** Dependence of polarization on pulse sequence delays. In all simulations, 100% enriched parahydrogen is assumed to be the initial state. **A.** The $\tau$ delay is optimized to maximize evolution under $J_{NH}$. When the resonances are within the pulse bandwidth ($\Delta\omega_{N,fb} = \Delta\omega_{N,f} - \Delta\omega_{N,b} = 0.9\ kHz$, blue), then a smooth response is obtained. Increasing $\Delta\omega_{N,fb}$ to approaching the edge of the pulse bandwidth introduces small artifacts that modulate the signal (red). **B.** The $\tau_{HH}$ delay shows highly oscillatory dynamics due to evolution under $\widehat{\mathcal{H}}_Z$. The four steps in the phase cycle, shown by the traces in color, have a modulus (black) that ensures polarization is obtained at any point of the delay.

It is interesting to examine the robustness of this experiment to the ligand exchange rate (**Fig. 4**). Decreasing the total length of the pulse sequence ($T = 8\tau + \tau_{HH} + 10t_p$) increases the performance at fast exchange rates, as more species experience the full pulse seuqence before exchange. To further minimize $T$, increasing $\omega_1 = 5\ kHz$ (90°) and $\omega_1 = 10\ kHz$ (180°) increases the average power dissipation to only $\langle\omega_1\rangle \approx 600\ Hz$ on both channels and improves polarization transfer for fast exchange. The broadband X-SABRE seuqence performs similarly to the ¹H dLIGHT experiment(21) for slow to intermediate exchange rates ($k_N \leq 200\ s^{-1}$). However, the broadband X-SABRE experiment outperforms the ¹H dLIGHT sequence at fast exchange rates likely due to the $t^3$ scaling (**c.f. eq. 13.3**) of magnetization build-up under this sequence compared to the $t^6$ scaling of the ¹H dLIGHT experiment. To make a fair and simple comparison, the parameters used in each pulse sequence were optimized under different experimental conditions and the only the enhancement under the optimal sequence at a given exchange rate is shown.

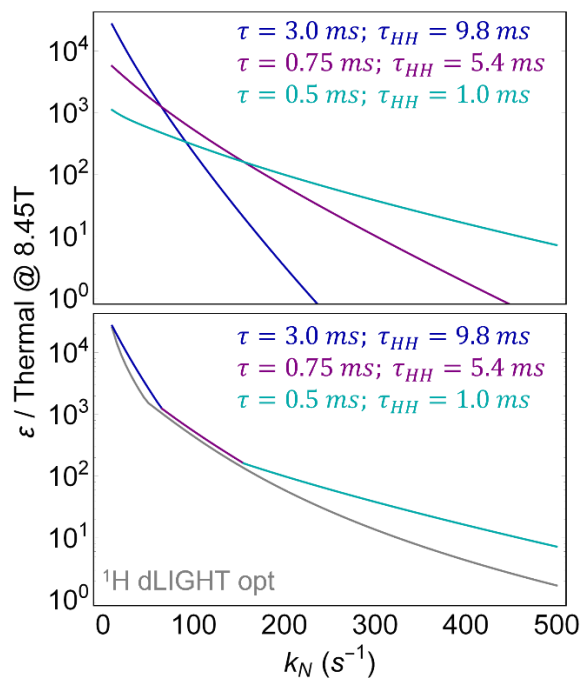

**Fig. 4** Robustness of broadband X-SABRE sequence to exchange rate in addition to increased bandwidth. The notable simulation parameters were $\omega_1 = 5\ kHz$ (90°), $\omega_1 = 10\ kHz$ (180°), $t_d = 315\ ms$, $k_H = 2\ s^{-1}$, and $[Ir]/[S] = 1/20$. All simulations were for 20 s. **Top.** Decreasing the length of the pulse sequence ($T$) permits larger signal enhancements at fast exchange rates and makes the sequence less sensitive to varying the exchange rate. **Bottom.** Comparing the best performance of the three broadband X-SABRE solutions (shown together) to the $^1$H decoupled LIGHT-SABRE ($^1$H dLIGHT) sequence (gray) indicates similar performance at low exchange rates but that the broadband sequence outperforms the $^1$H dLIGHT experiment at fast exchange rates ($k_N > 200\ s^{-1}$).

**Experimental**

All samples were prepared in methanol-d₄ (Cambridge Isotope) and activated under 43% enriched parahydrogen, which was bubbled through the sample with a head pressure of ~7 bar. The SABRE precatalyst was [Ir(IMes)(COD)]Cl (IMes = 1,3-bis(2,4,6-trimethylphenyl) imidazol-2-ylidene, COD = 1,5-cyclooctadiene) and was prepared at a concentration of ~3 mM for each sample (exact concentrations given in Figure captions). When hyperpolarization inactive (natural abundance nitrogen) coligands (C) are used to enhance exchange, complexes of the form [Ir(H)₂(IMes)(L)ₓ(C)₃₋ₓ]⁺ are generated, as was the case for the asymmetric SABRE system. This sample was prepared with L = ¹⁵N-acetonitrile (Cambridge Isotope, 98%) to a concentration and C = pyridine (natural abundance). Spectra were collected in medium wall high pressure NMR tubes (Wilmad) with a 5 mm OD. All broadband SABRE experiments were performed on a Bruker 360DX (8.45T) spectrometer operating 360 MHz (¹H) and 36.5 MHz (¹⁵N) using a 5 mm ¹H/²H/¹⁵N probe. Experiment-specific pulse sequence parameters are the same as those used in the simulations for **Fig. 3A**. All spectra were collected using a ²H-lock to prevent field drift. Bubbling of the parahydrogen was turned off 2-3 seconds prior to acquisition to reduce spectral inhomogeneities.

**Results and Discussion**

Hyperpolarization experiments are typically quantified using signal enhancement over thermal ($\varepsilon$) as well as polarization ($P$) units. However, these metrics do not reflect the actual magnitude of the NMR signal. The magnetization ($M$) of the sample is a critical metric, as nonlinear effects in NMR and next generation applications, like the CASPEr-Wind experiment proposed to search for axion-like dark matter(27), are proportional to the magnetization of the sample rather than the polarization. Here, we show that X-

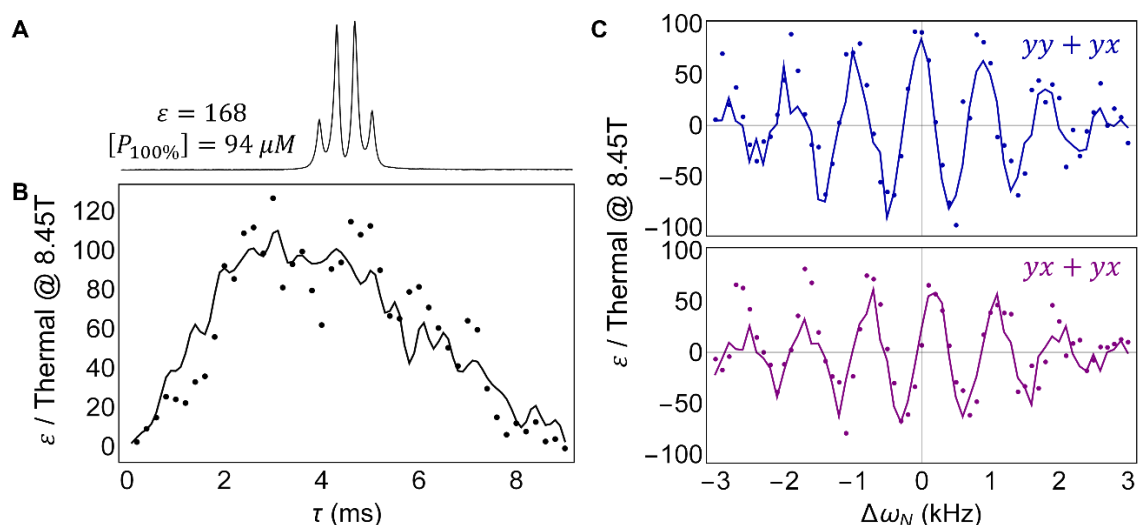

**Fig. 5** Experimental performance of broadband phase coherent SABRE sequence using $\tau_{HH} = 21\ ms$. All data were collected using ¹⁵N-acetonitrile (191 mM), pyridine (25 mM), and [Ir(IMes)(COD)]Cl (3.7 mM). Simulations (solid lines) assumed a target exchange rate of $k_N = 16\ s^{-1}$ and parahydrogen exchange rate of $k_{a,H} = 2\ s^{-1}$. **A.** Hyperpolarized spectrum with a signal enhancement of $\varepsilon = 168$ using the $yy \rightarrow yx$ excitation phase. This corresponds to a concentration of 100% polarized compound of $[P_{100\%}] = 94\ \mu M$. **B.** Scan of the $\tau$ delay of the pulse sequence with the optimum at $\tau = 3\ ms$. The excitation phase used for the pulse sequence was $yy \rightarrow yx$ and $\Delta\omega_N \approx (\Delta\omega_{N,b} + \Delta\omega_{N,f})/2$. **C.** Scan of the ¹⁵N resonance offset pulses using the $yy + yx$ (blue) and $yx + yx$ (purple) steps of the phase cycle.

SABRE hyperpolarization may be generated on $^{15}$N-acetonitrile by the proposed pulse sequence with $\varepsilon = 168$ (**Fig. 5A**). This corresponds to a molar polarization (concentration of 100% polarized targets) of $[P_{100\%}] = 94\ \mu M$, which is approximately twice the thermally polarized magnetization of neat $^{15}$N-acetonitile at this field ($[P_{100\%}] = 55\ \mu M$). The pulse sequence was optimized by scanning the $\tau$ delay, which was optimized at $\tau = 3\ ms$ (**Fig. 5B**). This delay is used for all of the experiments shown here. To optimize evolution under the $J_{NH}$ coupling, one would expect that the delay $\tau = 1/8J_{NH} \approx 5\ ms$ as this fully converts the density matrix into the desired coherence ($2\pi J_{NH} \times 4\tau = \pi/2$). However, the length of the pulse sequence grows as $8\tau$ and exchange attenuates performance at longer delays. The experimental data agrees well with the simulation when the target exchange rate is $k_N = 16\ s^{-1}$ and the parahydrogen replenishment rate of $k_H = 2\ s^{-1}$.

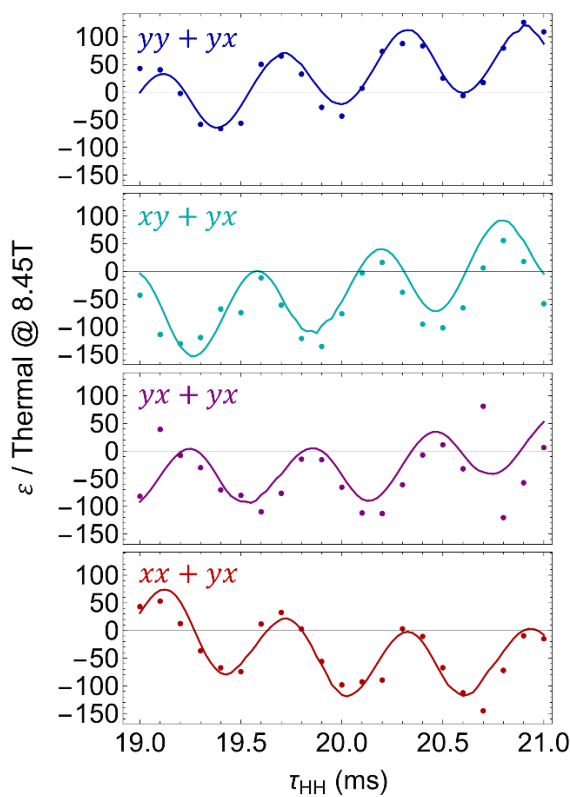

**Fig. 6** Hyperpolarization dynamics for the four-steps of the phase cycle required to remove dependence of $\widehat{\mathcal{H}}_Z$ when scanning the $\tau_{HH}$ delay. All data were collected using $^{15}$N-acetonitrile (191 mM), pyridine (25 mM), and [Ir(IMes)(COD)]Cl (3.7 mM). Simulations (solid lines) assumed a target exchange rate of $k_N = 16\ s^{-1}$ and parahydrogen exchange rate of $k_{a,H} = 2\ s^{-1}$. Performing the four-step phase cycle allows significantly lower sampling rates when scanning $\tau_{HH}$ and improving the efficiency of sequence optimization.

Hyperpolarization was observed over a 6 kHz bandwidth by scanning $\Delta \omega_N$, where evolution under $\widehat{\mathcal{H}}_Z$ can be compensated using the excitation phases $yy + yx$ and $yx + yx$ (**Fig. 5C**). The resonance offsets where the $yy$ excitation phase generates no signal correspond to a coherence that is $\pi/2$ out of phase from the $yx$ excitation phase of the second block meant to generate $\widehat{S}_z$. Shifting the excitation phase of the first step to $yx$ prepares the system in a state where the $\pi/2$ evolution under the $^{15}$N resonance offset subsequently brings the coherence into the correct phase to generate $\widehat{S}_z$. The hyperpolarization appears to oscillate approximately every 1 kHz. This is a result of aliasing in the data collection as it would be impractical to sample the data at a sufficient rate. The phase cycling circumvents needing prior knowledge of the precise value of the chemical shift of the iridium-bound heteronucleus. Finally, the excitation bandwidth is significantly larger than that of all previously reported methods for high field X-SABRE. This effectively eliminates the need to know specific molecular parameters such as the chemical shift.

Evolution under $\widehat{\mathcal{H}}_Z$ may be completely removed from the $\tau_{HH}$ delay using the four step phase cycle discussed previously (**Fig. 6**). Doing so eliminates the need to precisely optimize the $\tau_{HH}$ delay for hyperpolarization. Thus, this delay can be optimzied using much lower sampling rates than would be required with only one step of the phase cycle. Even the two-step cycle used to compensate for the $^{15}$N resonance offset would still generate dynamics that were modulated by evolution under the Zeeman terms for the $^1$H nuclei. It should be noted that the $\tau_{HH}$ delays chosen here ($\tau_{HH} = [19, 21]\ ms$) will generate large signals due to the slow exchange rate of $^{15}$N-acetonitrile ($k_N = 16\ s^{-1}$, extracted by simulation). For targets that exchange more rapdily, this would be too long of a period to efficiently generate polarization. In turn, the length of $\tau_{HH}$ would need to be decreased so that phase coherence was retained during the experiment. The excellent agreement between experiment and simulation for these cases indicates that much of the optimization may be done *in silico* before the experiment is performed.

This work permits simultaneous hyperpolarization of multiple components. The simplest extension of the previous expeirments is to hyperpolarize [15]N-acetonitrile and [15]N-pyridine simultaneoulsy (**Fig. 7**). In this case, all of the chemical shifts are known, and thus the excitation frequency can be chosen to be approximately central to these resonances. The chemical shift dispersion in this system is about 4 kHz at 8.45T, ranging from the iridum-bound [15]N-acetonitrile ($\delta \approx 120\ ppm$) resonance to the free [15]N-pyridine resonance ($\delta \approx 230\ ppm$), which is within the observable excitation bandwidth of the experiment under the parameters used here. Furthermore, there are only two dominant hyride resonances corresponding to the trans-([15]N-acetonitrile) and trans-([15]N-pyridine) hydrides, meaning there is only one value of $\Delta\omega_{HH}$. Thus, only one excitation frequency is required to generate sufficient hyperpolarization. For this case, $\tau_{HH} = 7.5\ ms$ to facilitate hyperpolarization with [15]N-pyridine, as it exchanges with the iridium catalyst at a rate of $k_N = 50\ s^{-1}$.

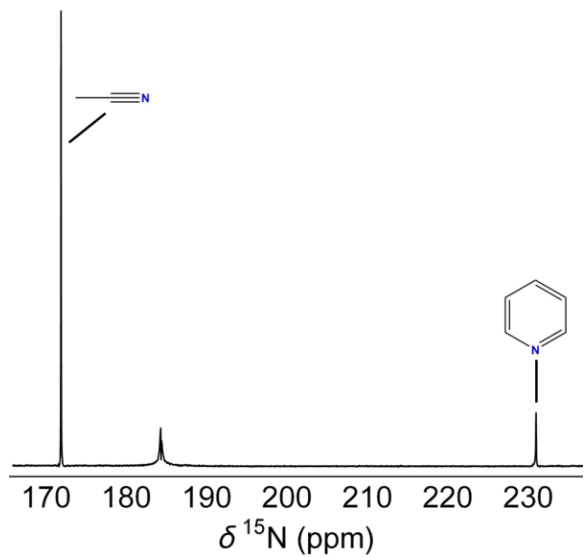

**Fig. 7** Simultaneous hyperpolarization of two [15]N-targets at high magnetic field. [15]N-acetonitrile (191 mM) and [15]N-pyridine (50 mM) can be co-polarized by irradiating between the free species resonances. The spectrum is the RMS of the four spectra collected at each step of the phase cycle. [1]H decoupling was utilized during the FID acquisition to remove *J*-couplings from the spectrum. The unlabeled resonance is that of iridium-bound [15]N-pyridine. For these experiments, $\tau_{HH} = 7.5\ ms$.

This expeirment is capable of hyperpolarizing systems where the chemical shift of the iridium-bound species is unobservable, such as in samples with multiple componenets. The concentration of each iridium-bound species decreases as the number of components in the SABRE sample grows, making it difficult to observe the iridium-bound resonance in the spectrum. Here, we demonstrate simultaneous hyperpolarization on [15]N-pyridine, [15]N-nicotinamide, [15]N-benzylamine, and [15]N$_2$-imidazole. In this system, the chemical shift dispersion of just the free species is on the order of the total dispersion of the [15]N-acetonitrile/[15]N-pyridine SABRE system. While the pulse bandwidth could be increased, this would lead to higher power deposition in the sample. Instead, the pulse carrier frequency can be swept across the spectrum to excite different regions (**Fig. 8A**), where a

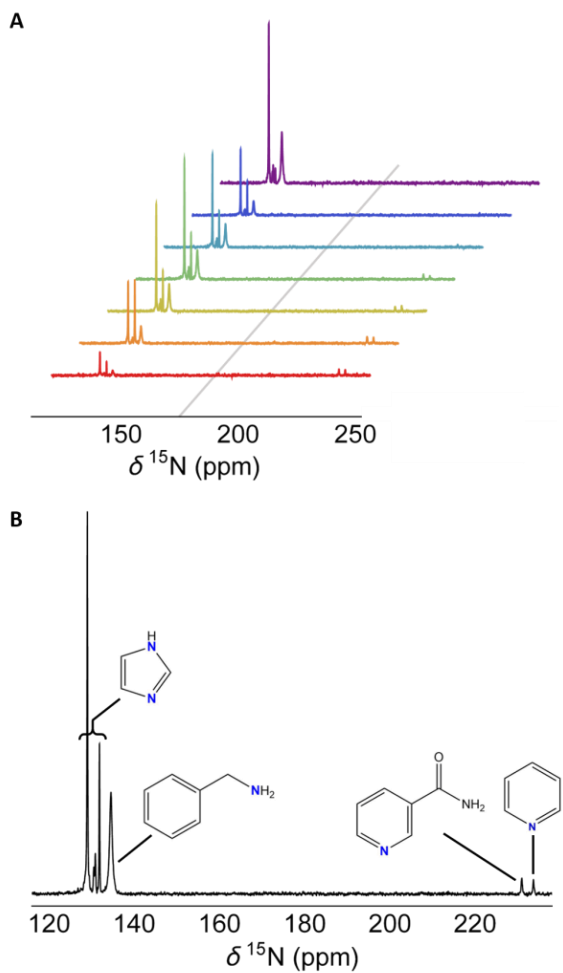

**Fig. 8** Simultaneous hyperpolarization of multiple [15]N-targets at high magnetic field. [15]N-pyridine (53 mM), [15]N-nicotinamide (59 mM), [15]N-benzylamine (125 mM), and [15]N$_2$-imidazole can be co-polarized using 5 mM [Ir(IMes)(COD)]Cl using $\tau_{HH} = 7.5\ ms$. **A.** Scanning a 7 kHz bandwidth in 1 kHz increments across the [15]N spectrum highlights the regions where different species are hyperpolarized. Each of these is a root-mean-square (RMS) of the spectra from the four-step phase cycle. There are multiple spectra where all of the species are polarized but with reduced signal. The gray line is shown as a guide for the eye. **B.** Co-adding the resulting spectra increases all of the resulting signals and can be utilized to effectively increase the bandwidth without increasing power deposition during the pulse sequence.

7 kHz bandwidth was excited in 1 kHz increments. Sampling at this rate was not required, but clearly indicates where the downfield resonances corresponding to $^{15}$N-pyridine and $^{15}$N-nicotinamide ($\delta \approx 230\ ppm$) are no longer polarized by the pulse sequence. Then, the resulting spectra may be added to increase the signal-to-noise ratio of the spectra (**Fig. 8B**). Using this method, the entire $^{15}$N-spectral window could be rapidly hyperpolarized, making X-SABRE hyperpolarization of complex systems feasible for the first time. Importantly, none of the chemical shifts of the bound or free species had to be known *a priori* for these experiments to work.

## Conclusions

We have developed a new broadband, phase coherent excitation strategy for generating hyperpolarized X-SABRE magnetization at high field. By phase cycling the excitation phase of the X-SABRE hyperpolarization, this technique is insensitive to resonance offset of the iridium-bound and free target. In addition to the observed 6 kHz hyperpolarization bandwidth in these experiments, this affords the ability to simultaneously hyperpolarize multiple targets with X-SABRE for the first time. Doing so provides access to new applications for hyperpolarized magnetic resonance, such as SABRE hyperpolarized metabonomics and biomolecular NMR. Additionally, the presence of quadrupolar spins ($^2$H, $^{14}$N) in targets will not detrimentally affect the hyperpolarized signal at high field, as these nuclei do not act as relaxation sinks under these conditions.

## Conflicts of interest

There are no conflicts to declare.

## Acknowledgements

This work was supported by the National Science Foundation grants CHE-1665090 and CHE-2003109.

## Notes and references

# Supplementary Information: Phase coherent excitation of SABRE permits simultaneous hyperpolarization of multiple targets at high magnetic field

Jacob R. Lindale[a], Shannon L. Eriksson[a,b], Warren S. Warren[c]

[a] Department of Chemistry, Duke University, Durham, NC, 27708. [b] School of Medicine, Duke University, Durham, NC, 27708. [c] Departments of Physics, Biomedical Engineering, and Radiology, Duke University, Durham, NC, 27708.

## ¹H decoupled LIGHT-SABRE simulations

The broadband X-SABRE experiment introduced here is only the second experiment to generate hyperpolarization in asymmetric ligand environments at high field. Comparing its performance to the ¹H decoupled LIGHT-SABRE (¹H dLIGHT) experiment(1) (**Fig. S1**) provides insight on the various benefits of this method. The pulse sequence parameter space for the ¹H dLIGHT experiment spans several parameters, such as the SLIC-pulse length $t_p$, the exchange delay $t_d$, the SLIC pulse power $\omega_{1,N}$ and resonance offset $\Delta\omega_N$, and the ¹H decoupling power $\omega_{1,H}$. As both of these experiments were benchmarked on the ¹⁵N-acetonitrile/¹⁴N-pyridine X-SABRE complex, we will use simulations of this system also as a reference. We will fix ¹H decoupling power at the experimentally optimized $\omega_{1,H} = 1.6\ kHz$.

This phase space may be efficiently explored *in silico* with the DMExFR2(2) simulation package built and experimentally verified for SABRE. Broadly, we find that hyperpolarization is optimized for $t_d = 0\ ms$, i.e. continuously irradiating on both ¹H and ¹⁵N at the respective conditions. This significantly reduces the size of the parameter phase space that must be searched to $\omega_{1,N}$ and $\Delta\omega_N$, which is shown in **Fig. S2** for $k_N = 16\ s^{-1}$ and $k_N = 150\ s^{-1}$. The position of the optimum is marked by the solid crosshairs and the position of the level anti-crossing (LAC) predicted optimum is shown with the dashed crosshair. These pulse sequence parameters were used to generate the exchange rate dependent simulations shown in **Fig. 4** in the main text.

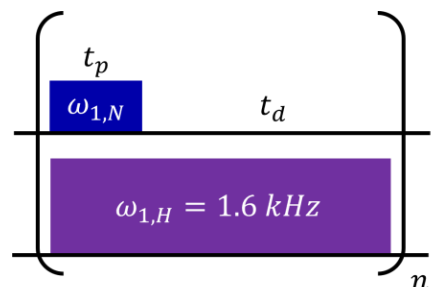

**Fig. S1** ¹H decoupled LIGHT-SABRE experiment. Continuous irradiation on ¹H preserves the $\hat{I}_1 \cdot \hat{I}_2$ spin order and permits SLIC-based hyperpolarization methods in asymmetric ligand environments. For ¹⁵N-acetonitrile at 8.45T, $\omega_{1,H} = 1.6\ kHz$ was experimentally optimized and is used in all of the following simulations.

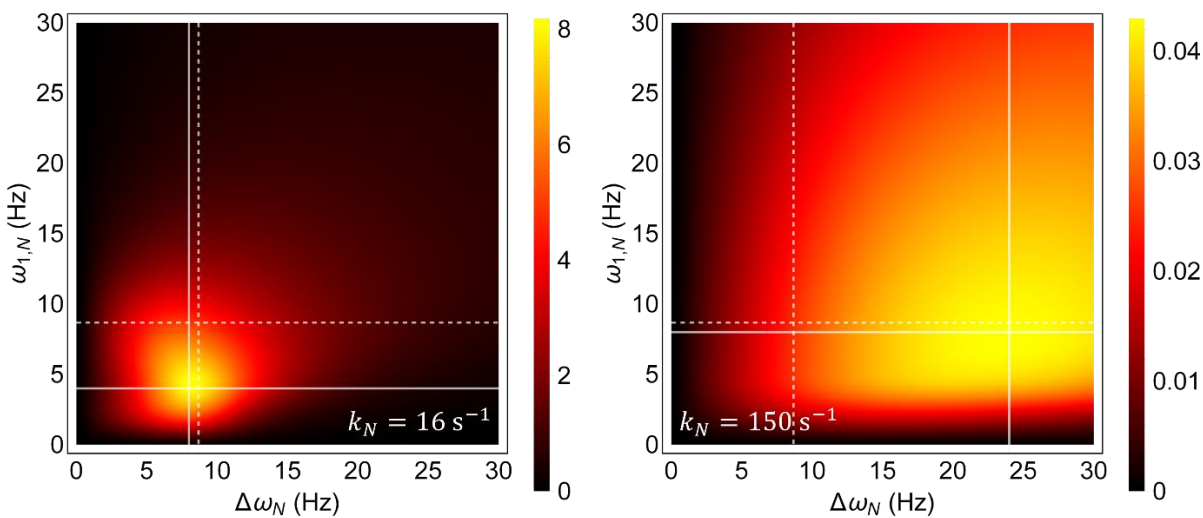

**Fig S2** Optimization of ¹H decoupled LIGHT-SABRE simulations for ¹⁵N-acetonitrile at 8.45T. For these simulations, $k_H = 2\ s^{-1}$, $\omega_{1,N} = 1.6\ kHz$, and $[Ir]/[S] = 1/20$. All simulations were run for 20s. The positions of the optimum conditions are indicated by solid crosshairs. The dashed crosshairs indicate the condition predicted by the level anti-crossing, which is $\Delta\omega_N \approx \omega_{1,N} \approx J_{HH}$. Notably, the optimal hyperpolarization conditions are found at positions that are nearly a factor of two different from the LAC prediction for both cases.

## Refocusing the $\tau_{HH}$ delay

The broadband X-SABRE experiment utilizes phase cycling to remove the effects of $\hat{\mathcal{H}}_Z$, which manifests as oscillatory structures in both $\Delta\omega_N$ and $\tau_{HH}$. A pair of $180_y$ pulses (or other phases) may be added to the pulse sequence at $\tau_{HH}/2$ to refocus evolution under $\hat{\mathcal{H}}_Z$, which theoretically should greatly simplify the experiment and remove the need for phase cycling (**Fig. S3A**). However, this changes the net flip angle of the experiment to $(2n+1)\pi$ and has the effect that it inverts the hyperpolarization with each successive application of the pulse sequence. Refocusing the $\tau_{HH}$ delay does, in fact, remove the oscillatory structure from the $\Delta\omega_N$ and $\tau_{HH}$ profiles, but greatly reduces the resultant hyperpolarization under this pulse sequence (**Fig. S3B**).

Notably, evolution under $\hat{\mathcal{H}}_Z$ will have an adverse effect when hyperpolarizing in strongly inhomogeneous fields, as phase cycling will not be able to compensate for the inhomogeneities. However, the experiments here were performed with continuous parahydrogen bubbling at 8.45T, where the large susceptibility difference between the gaseous and liquid phases generates large inhomogeneities which are most prominent on the $^1$H channel. Even still, hyperpolarization was readily achieved with this pulse sequence.

## Simulation parameters

The $^{15}$N-acetonitrile system was characterized at 8.45T, and the key parameters for this system are given here in detail. The Hamiltonian of this system is:

$$\hat{\mathcal{H}} = \Delta\omega_{HH}(\hat{I}_{1z} - \hat{I}_{2z}) + \omega_{Hm}\sum_{i=1}^{3}\hat{I}_{iz}$$
$$+ 2\pi(J_{HH}\hat{I}_1 \cdot \hat{I}_2 + J_{NH}\hat{I}_{1z}\hat{S}_z) \qquad (S1)$$
$$+ 2\pi J_{NHm}\sum_{i=1}^{3}\hat{I}_{iz}\hat{S}_z$$

The parameters $\omega_{Hm}$ and $J_{NHm}$ are the precession frequencies of the methyl $^1$H and the J-coupling to these nuclei from the $^{15}$N spin. We have assumed the same rotating frame as used in eq. 2 with $\Delta\omega_N = 0$ and have dropped the pulses for brevity. The numerical values of these parameters are:

$$\Delta\omega_{HH} = 460\ Hz \qquad (S2a)$$

$$\omega_{Hm} = 10\ kHz \qquad (S2b)$$

$$J_{NH} = -25.41\ Hz \qquad (S2c)$$

$$J_{HH} = -8.68\ Hz \qquad (S2d)$$

$$J_{NHm} = -1.69\ Hz \qquad (S2e)$$

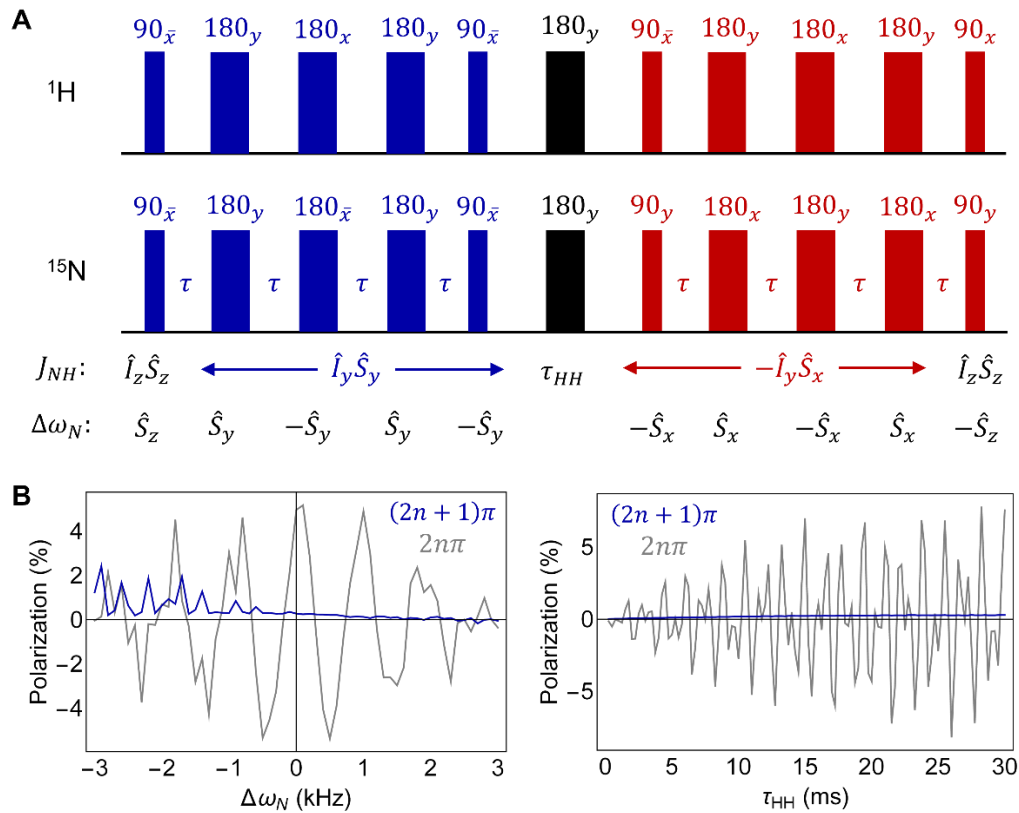

**Fig. S3** $\tau_{HH}$ refocused version of the broadband X-SABRE experiment. **A.** Evolution under $\hat{\mathcal{H}}_Z$ may be refocused by adding two $180_y$ pulses at $\tau_{HH}/2$. The significant difference in this sequence compared to the non-refocused version is that the net flip angle of this sequence is $(2n+1)\pi$ whereas the non-refocused version has a net flip angle of $2n\pi$. **B.** While the dependence on the resonance offset and $\tau_{HH}$ no longer have oscillatory structures due to the refocusing of $\hat{\mathcal{H}}_Z$, the resultant hyperpolarization is significantly lower over the entire pulse sequence. For the $\Delta\omega_N$ profile, $\tau_{HH} = 21\ ms$ and for the $\tau_{HH}$ scan, $\Delta\omega_N = 0\ Hz$. For all simulations, $\tau = 3\ ms$, $k_N = 16\ s^{-1}$, $k_H = 2\ s^{-1}$, and $[Ir]/[S] = 1/20$.

For the simulations fit to the experimental data, there was a $\Delta\omega_H = -734\ Hz$ and $\Delta\omega_{N,b} = 935\ Hz$ resonance offset for the bound species and $\Delta\omega_{N,f} = -857$ offset for the free $^{15}N$ species.